\documentclass[aps,amsmath,amssymb,prd,floatfix,reprint,superscriptaddress]{revtex4-2}
\usepackage{siunitx}
\usepackage{tikz}
\usepackage{bm}
\usepackage{longtable}
\usepackage{verbatim}
\usepackage{dcolumn}
\usepackage{graphicx}
\usepackage{listings}

\usepackage[version=4]{mhchem}
\usepackage{amsmath}
\usepackage{multirow}
\usepackage{hyperref}
\hypersetup{
	bookmarks=true,
	colorlinks=true,
	allcolors=blue
}

\newcommand{\jpsi}{J/\psi}
\newcommand{\piz}{\pi^{0}}

\newcommand{\az}{a_0(980)}
\newcommand{\bo}{b_1(1235)^0}
\newcommand{\fz}{f_0(980)}
 \newcommand{\BESIIIorcid}[1]{\href{https://orcid.org/#1}{\hspace*{0.1em}\raisebox{-0.45ex}{\includegraphics[width=1em]{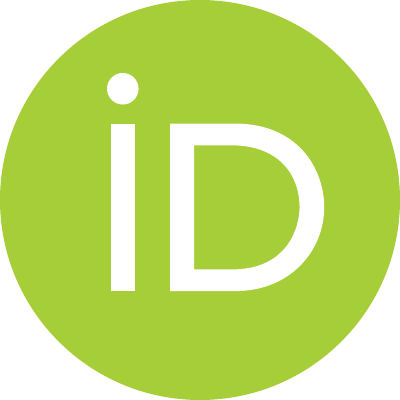}}}}
\DeclareSIUnit\clight{\text{\ensuremath{c}}}

\begin{document}


\title{\boldmath Amplitude Analysis of the Isospin-Violating Decay $\jpsi\to\gamma\eta\piz$}

\author{
M.~Ablikim$^{1}$\BESIIIorcid{0000-0002-3935-619X},
M.~N.~Achasov$^{4,c}$\BESIIIorcid{0000-0002-9400-8622},
P.~Adlarson$^{81}$\BESIIIorcid{0000-0001-6280-3851},
X.~C.~Ai$^{87}$\BESIIIorcid{0000-0003-3856-2415},
C.~S.~Akondi$^{31A,31B}$\BESIIIorcid{0000-0001-6303-5217},
R.~Aliberti$^{39}$\BESIIIorcid{0000-0003-3500-4012},
A.~Amoroso$^{80A,80C}$\BESIIIorcid{0000-0002-3095-8610},
Q.~An$^{77,64,\dagger}$,
Y.~H.~An$^{87}$\BESIIIorcid{0009-0008-3419-0849},
Y.~Bai$^{62}$\BESIIIorcid{0000-0001-6593-5665}
% ...
\\
\vspace{0.2cm}
(BESIII Collaboration)\\
\vspace{0.2cm} {\it
$^{1}$ Institute of High Energy Physics, Beijing 100049, People's Republic of China\\
% ...
$^{4}$ Budker Institute of Nuclear Physics SB RAS (BINP), Novosibirsk 630090, Russia\\
% ...
$^{31}$ INFN Sezione di Ferrara, (A)INFN Sezione di Ferrara, I-44122, Ferrara, Italy; (B)University of Ferrara, I-44122, Ferrara, Italy\\
% ...
$^{39}$ Johannes Gutenberg University of Mainz, Johann-Joachim-Becher-Weg 45, D-55099 Mainz, Germany\\
% ...
$^{62}$ Southeast University, Nanjing 211100, People's Republic of China\\
% ...
$^{64}$ State Key Laboratory of Particle Detection and Electronics, Beijing 100049, Hefei 230026, People's Republic of China\\
% ...
$^{77}$ University of Science and Technology of China, Hefei 230026, People's Republic of China\\
% ...
$^{80}$ University of Turin and INFN, (A)University of Turin, I-10125, Turin, Italy; (B)University of Eastern Piedmont, I-15121, Alessandria, Italy; (C)INFN, I-10125, Turin, Italy\\
$^{81}$ Uppsala University, Box 516, SE-75120 Uppsala, Sweden\\
% ...
$^{87}$ Zhengzhou University, Zhengzhou 450001, People's Republic of China\\
\vspace{0.2cm}
$^{\dagger}$ Deceased\\
% ...
$^{c}$ Also at the Novosibirsk State University, Novosibirsk, 630090, Russia\\
% ...
	}
}


\date{\today}

\begin{abstract}
	    Using $(10\,087 \pm 44) \times 10^6$ $J/\psi$ events collected with the BESIII detector, we perform the first amplitude analysis of the process $J/\psi\to\gamma\eta\pi^0$. The decay is dominated by the intermediate processes $J/\psi\to\pi^0 b_1(1235)^0 \to\gamma\eta\pi^0$, $J/\psi\to\pi^{0}\rho(1450)^0 \to\gamma\eta \pi^0$ and $J/\psi\to\eta h_1(1170) \to\gamma\eta\pi^0$.
    Contributions from $J/\psi\to\gamma a_0(980)^0\to\gamma\eta\pi^0$, $J/\psi\to\gamma a_2(1320)^0\to\gamma\eta\pi^0$ and $J/\psi\to\gamma
	a_2(1700)^0\to\gamma\eta\pi^0$ are observed with a statistical significance exceeding $5\sigma$, constituting the first observation of radiative transitions
of $J/\psi$ to isospin-triplet scalar mesons. The total branching fraction of $J/\psi\to\gamma \eta \pi^0$ is measured to be $(25.7\pm0.3\pm1.5)\times 10^{-6}$, where the first uncertainty is statistical and the second systematic. This result is consistent with the previous measurement, with the precision improved by more than a factor of two.

\end{abstract}
\maketitle

\section{Introduction}%
\label{sec:introduction}

The radiative decay $\jpsi\to\gamma X$ provides a clean environment in which to search for glueballs~\cite{Reinders:1984sr,*Morningstar:1999rf,*Gross:2022hyw,BESIII:2023wfi} and hybrid mesons~\cite{BESIII:2022riz}. The dominant radiative decay of the
$\jpsi$ proceeds via $\jpsi\to\gamma gg$, where $g$ denotes a gluon. The formation of isospin-triplet final states from a two gluon system is highly suppressed by isospin conservation, and radiative $\jpsi$ decays to isospin-triplet states are expected to be strongly suppressed. Nevertheless, sizable enhancements may arise through isospin-symmetry-breaking effects or new interaction mechanisms~\cite{Kiselev:2008zzb}. Consequently, studies of radiative $\jpsi$ decays to isospin-triplet states offer a sensitive test of isospin conservation and of possible new dynamics.

Among isospin-multiplet mesons, the $\az$ is the lightest scalar. 
Theoretical work suggests that it is not a conventional $q\bar{q}$ state~\cite{Baru:2003qq,*Baru:2004xg,*Baru:2010ww}, a picture supported by its production in $\phi\to\gamma\eta\piz$~\cite{Achasov:2000ku,*KLOE:2002kzf}
and by $\az$-$\fz$ mixing in $\jpsi\to\phi\eta\piz$~\cite{BESIII:2010dhc,
*BESIII:2018ozj,*BESIII:2023zwx}. The decay $\jpsi\to\gamma\az^0$ offers new insights into the nature of $\az$. Its branching fraction ($\mathrm{BF}$) is expected to be an order
of magnitude smaller than that of $\jpsi\to \phi(\omega) \eta\piz$, making it highly sensitive to exotic production mechanisms~\cite{Sakai:2019uig, Xiao:2019lrj}.
Predictions for $\jpsi\to\gamma\az^0$ vary widely among phenomenological models in quantum chromodynamics (QCD). Within the vector-meson-dominance (VMD) framework detailed in Ref.~\cite{Sakai:2019uig}, the $\az$ is treated as a dynamically generated state arising from final-state interactions, whereas Ref.~\cite{Xiao:2019lrj} incorporates $\fz$-$\az$ mixing and obtains a significantly different BF, as shown later in Table~\ref{tab:a0}.  

The unflavored axial-vector meson $\bo$ ($J^{PC}=1^{+-}$), discovered six 
decades ago~\cite{Baltay:1967zza}, remains poorly understood~\cite{10.1093/ptep/ptaa104}. Its internal structure~\cite{Clymton:2023txd,*Xie:2023cej} and
decay properties are debated; several studies interpret it as a molecular state generated by meson–meson interactions~\cite{Lutz:2003fm,*Roca:2005nm,*Liang:2019vhf} rather than as a conventional $q\bar{q}$ configuration. A striking example is its radiative decay. The only measured $b_1(1235)$ radiative mode, $b_1(1235)^+\to\gamma\pi^+$, has a partial width of $(230\pm60)~\mathrm{keV}$~\cite{Collick:1984dkp}, far exceeding quark-model predictions of ${(66-184)}{~\rm keV}$~\cite{Rosner:1980ek,*Ishida:1988uw,*Aznaurian:1990kk,Lutz:2008km,Jeong:2018exh}. This discrepancy has motivated alternative explanations, including meson–meson interaction models~\cite{Roca:2006am, Lutz:2008km} and a pentaquark condensate~\cite{Jeong:2018exh}. An independent measurement of $b_1(1235)^0\to\gamma\eta$ is therefore essential. Theoretical predictions for this mode span a wide range (Table~\ref{tab:b1}), with Ref.~\cite{Nagahiro:2008zza} favoring tree-level VMD dominance and Ref.~\cite{Lutz:2008km} emphasizing loop-level $K^*\bar{K}$ re-scattering.  
Both interpretations assume a dynamically generated $b_1(1235)^0$, and a first experimental determination will provide a decisive test.  
Similar questions motivate studies of the $h_1$ family~\cite{Dankowych:1981ks,BNL-E852:2000poa}, which are expected to share decay mechanisms with $b_1(1235)$~\cite{Nagahiro:2008zza,Clymton:2024pql}.

The decay $\jpsi\to\gamma\eta\piz$ is an ideal laboratory for investigating isospin-violating decays and the properties of $b_1^0$ mesons.  
Its isospin-violating nature suppresses gluonic backgrounds ($\jpsi\to\gamma gg$), providing clean access to radiative transitions such as $\jpsi\to\gamma a_{0,2}^0$ and $b_1^0(h_1)\to\gamma\eta(\piz)$.  
A previous BESIII study~\cite{BESIII:2016gkg}, limited by statistics, obtained $\mathrm{BF}(\jpsi\to\gamma\eta\piz)=\num{2.14\pm0.18\pm0.25e-5}$ and set 90\,\% confidence-level upper limits of $\num{2.5e-6}$ and $\num{6.6e-6}$ on $\mathrm{BF}(\jpsi\to\gamma a_0(980)^0)$ and $\mathrm{BF}(\jpsi\to\gamma a_2(1320)^0)$, respectively.  
In this paper, we report the first amplitude analysis of $\jpsi\to\gamma\eta\piz$, based on the unprecedented BESIII data set of \num{10087 \pm 44 e6} $\jpsi$ events~\cite{BESIII:2021cxx}, yielding greatly improved precision and enabling the isolation of individual intermediate states.

\section{Event Selection}%
\label{sec:event_selection}
A detailed description of the design and performance of the \mbox{BESIII} detector, and the simulated data samples can be found in the supplemental
material~\cite{supplementalMaterial}.
The $\eta$ and $\piz$ mesons are reconstructed via  $\eta\to\gamma\gamma$ and $\piz\to\gamma\gamma$. Signal candidates are required to have at least five photon candidates and no charged tracks.
A four-constraint (4C) kinematic fit imposing overall energy-momentum conservation is performed on all five photon combinations, and the one with the smallest
$\chi^2_{4C}$ ($<25$) is retained.
The resulting four-momenta are used for subsequent background suppression. 
The $\piz$ and $\eta$ candidates are chosen from the five photons by selecting the combination whose two-photon invariant masses best match the nominal $\piz$ and $\eta$ masses~\cite{10.1093/ptep/ptaa104}, using the corresponding mass resolutions of \SI{5}{MeV/\clight^2} and \SI{9}{MeV/\clight^2}.
A six-constraint (6C) kinematic fit that in addition constraining the $\piz$ and $\eta$ invariant masses to their known values~\cite{10.1093/ptep/ptaa104} is then applied to improve the resolution, and the resulting four-momenta are used as input to the amplitude analysis. 

Inclusive MC studies~\cite{ZHOU2021107540} show that the dominant backgrounds arise from incorrect photon pairings in $\piz$ and $\eta$ reconstruction.
To mitigate these backgrounds, a discriminating variable $\Delta^2_{\piz}$ is introduced:
	$\Delta^2_{\piz}=(m_{\gamma_1\gamma_2}-m_{\piz}^{\mathrm{PDG}})^2+(m_{\gamma_3\gamma_4}-m_{\piz}^{\mathrm{PDG}})^2$.
After evaluating all possible four-photon combinations, the requirement $\min\left(\Delta^2_{\piz}\right)> \SI{0.05}{GeV^2/\clight^4}$ is
imposed~\cite{BESIII:2016gkg}, which removes \SI{99.5}{\percent} of the combinatorial backgrounds while maintaining \SI{68}{\percent} of the signal events.
The following mass-window requirements are imposed to suppress backgrounds from $\omega\to\gamma\piz$ in the decays
$\jpsi\to\gamma\eta^{\prime}\to\gamma\gamma\omega$ and $\jpsi\to\piz
b_1(1235)^0\to\omega\piz\piz$:
\begin{itemize}
	\item $\left|m_{\pi^0\gamma_\eta} - m_\omega^{\mathrm{PDG}}\right| > 65$ MeV/$c^2$,
   \item $\left|m_{\gamma\gamma_\eta} - m_\omega^{\mathrm{PDG}}\right| > 65$ MeV/$c^2$,
\end{itemize}
and the following one for $\jpsi\to\gamma\eta\eta$ decay:
\begin{itemize}
\item $\left|m_{\gamma\gamma_{\pi^0}} - m_\eta^{\mathrm{PDG}}\right| > 39$ MeV/$c^2$,
\end{itemize}
where $\gamma_\eta$ ($\gamma_{\piz}$) is one of the photons from the selected $\eta$ ($\piz$) candidate,  
and $m_{\omega}^{\mathrm{PDG}}$ is the known mass of the $\omega$ meson~\cite{10.1093/ptep/ptaa104}. 
After these criteria, the MC studies indicate that the remaining background is dominated by $\jpsi \to \gamma\eta^\prime$ with subsequent decay $\eta^\prime
\to \gamma\gamma\piz$ or $\piz\piz\eta$, and $\jpsi \to \eta\omega$ or $\jpsi \to \eta\phi$ process with $\omega(\phi) \to \gamma\piz$.
Therefore, the events fulfilling $m_{\eta\piz}<\SI{1.0}{GeV/\clight^2}$ or $m_{\gamma\piz}< \SI{1.1}{GeV/\clight^2}$ are rejected.
Notably, these two requirements will suppress over \SI{90}{\percent} of the $\az^0$ signal in the $\eta\piz$ invariant mass distribution. 
A dedicated procedure, described in Sec.~\ref{sec:amplitude_analysis}, is used to estimate the signal yield in the region $m_{\eta\piz}<1.0~\mathrm{GeV}/c^{2}$. 

\section{Background Studies and Signal Extraction}
\label{sec:backgroundstudy}
Background contributions are classified into two categories:  
(I) those in which the $\eta$ and $\piz$ are correctly reconstructed, and  
(II) those in which at least one of them is mis-reconstructed.

Since the decay $\jpsi\to \eta \piz\piz$ is forbidden by the charge conjugation symmetry, the category~I background is dominated by the process
$\jpsi\to\gamma \eta \piz\piz$ with high $\eta \piz\piz$ mass, or from $\eta_c$ decays, where the radiative photon from $\jpsi$ or one of
the photons from a $\piz$ is too soft to be detected.  

The category-I background yield is estimated to be \num{225\pm15} using a data-driven method based on $J/\psi \to \gamma\eta\pi^0\pi^0$ candidates in the
$\eta_c$ mass region and phase-space (PHSP) MC simulation. The MC is corrected for the known $\eta_c$ radiative lineshape~\cite{CLEO:2008pln} and reweighted to reproduce the observed Dalitz-plot distribution before evaluating the yield passing the nominal selection.

Category-II backgrounds, arising mainly from mis-reconstructed $\eta$ and/or $\piz$ candidates, are separated from the signal with a $Q$-weight method~\cite{Williams:2008sh}, widely used in previous experiments~\cite{CLAS:2009hpc,*CrystalBarrel:2014tvm,*BESIII:2019hek,*BaBar:2023kug,*BESIII:2024ein}.  
In the $Q$-weight method, the distance between two events $i$ and $j$ is defined as 
$d_{i,j}^{2}=\sum_{k=1}^{5} \left[ \left(\xi_{k}^{i}-\xi_{k}^{j}\right)/\Delta_k \right]^2$,where $\vec{\xi}=(m^2_{\eta\piz},m^{2}_{\gamma\piz},m^{2}_{\gamma\eta},\theta_{\gamma},\theta_{\piz})$ contains the squared invariant masses of the $\piz\eta$, $\gamma\piz$ and $\gamma\eta$ combinations, along with the polar angles of $\gamma$ and $\piz$ in the lab frame. 
The normalization factors $\Delta_k$ are set to $\Delta_k=\max(\xi_k)-\min(\xi_k)$ to constrain the distance within $[0,1]$, with values
\SI{2.1}{GeV^2/\clight^4}, \SI{1.6}{GeV^2/\clight^4}, \SI{2.0}{GeV^2/\clight^4}, \SI{0.6}{rad} and \SI{0.6}{rad} for $m^2_{\eta\piz}$, $m^{2}_{\gamma\piz}$, $m^{2}_{\gamma\eta}$, $\theta_{\gamma}$ and $\theta_{\piz}$, respectively.
For each event, the 200 nearest neighbors are used in a two-dimensional unbinned maximum-likelihood fit to the $m_{\piz}$ versus $m_{\eta}$ distribution with probability density function (PDF):
\begin{align}
F_{\mathrm{total}} & = N_{1} \cdot (s_{\piz}\times s_{\eta}) + N_{2} \cdot (s_{\piz}\times b_{\eta}) \notag\\
                   & + N_{3} \cdot (b_{\piz}\times s_{\eta}) 
                     + N_{4} \cdot (b_{\piz}\times b_{\eta}),
\label{eq:fpdf}
\end{align}
\noindent
where $s_{\piz}$ ($s_{\eta}$ is the detected line shape for the $\piz$ ($\eta$) modeled with a Novosibirsk function~\cite{Belle:1999bhb}, and $b_{\piz}$
($b_{\eta}$) is the corresponding background function described by a second-order polynomial.  
The signal weight $Q_{i}$ for a certain event is then calculated as the ratio of  the signal  yield to the total yield within the region $\left|
m_{\piz}-m_{\piz}^{\mathrm{PDG}}\right|< \SI{15}{MeV/\clight^2}$ and $ \left| m_{\eta} - m_{\eta}^{\mathrm{PDG}}\right|< \SI{26}{MeV/\clight^2}$. 

\begin{figure}[htpb]
	\centering
	\includegraphics[width=0.8\linewidth]{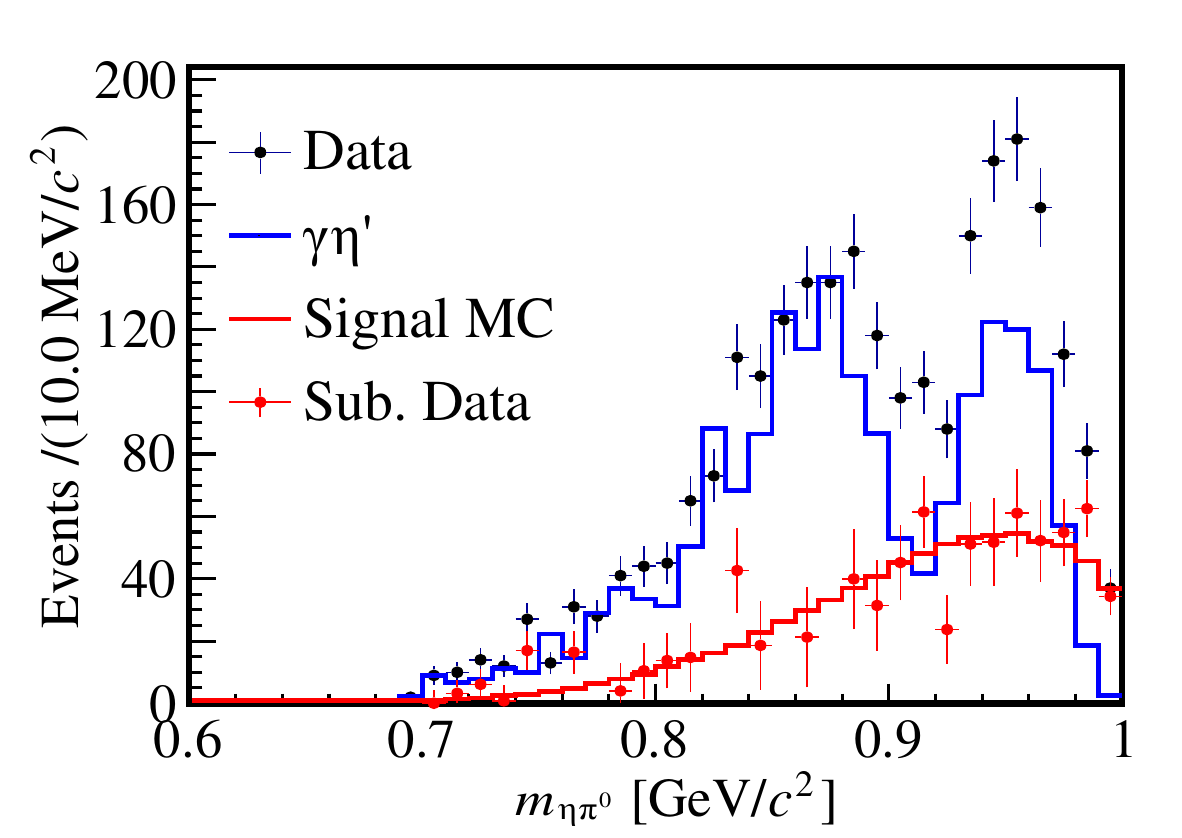}
	\caption{The $m_{\piz\eta}$ distribution at $m_{\piz\eta}< \SI{1.0}{GeV/\clight^2}$, where the black points with error bars correspond to data, the blue line represents the estimated background events, the red points with error bars represent the background-subtracted data, and the red line corresponds to the signal MC curve.}%
	\label{fig:mpieta_data}
\end{figure}

For the region $m_{\piz\eta}< \SI{1.0}{GeV/\clight^2}$, dominated by background, the signal yield is extracted with a side-band approach.
The detailed studies based on the inclusive MC sample and data indicate the background is dominantly from $\jpsi\to\gamma\eta^\prime$ with $\eta^\prime\to\gamma\gamma\piz$ or $\piz\piz\eta$, both of which produce smooth distributions in the $\eta\to\gamma\gamma$ mass spectrum.
The $m_{\eta\piz}$ distribution in the $\eta$ side-band region, defined as $40<|m_{\eta}-m_{\eta}^{\mathrm{PDG}}|<53~\mathrm{MeV}/c^{2}$, is fitted with a PDF
consisting of two background components. Their shapes are taken from MC simulation and convolved with a Gaussian function to account for the difference in mass resolution
between data and MC. The MC samples are generated according to the $\eta'$ decay kinematics described in Refs.~\cite{BESIII:2016oet, BESIII:2019gef}. The fitted background yields are extrapolated to the signal region, giving a net signal of $N_0=\num{676\pm80}$ events after background subtraction. 
Figure~\ref{fig:mpieta_data} shows the $m_{\eta\piz}$ distribution of the data sample in the $\eta$ signal region, as well as that of the simulated signal and background events.
The signal MC shape is based on the amplitude analysis model described in Sec.~\ref{sec:amplitude_analysis}.
The background-subtracted data agree well with signal MC, indicating that the $m_{\eta\piz}<1.0~\mathrm{GeV}/c^{2}$ region is dominated by $\az^0$ and that the subtraction does not distort the lineshape.

\section{Amplitude Analysis}%
\label{sec:amplitude_analysis}
We perform the first amplitude analysis of $\jpsi\to\gamma\eta\piz$ to determine the branching fractions of intermediate states and search for additional resonances.
 The analysis is performed by using the \texttt{GPUPWA} framework~\cite{Berger:2010zza}, which constructs the quasi-two-body decay amplitudes for the three sequential processes $\jpsi\to\gamma\ X (\to\eta\piz)$, $\jpsi\to\piz\ Y (\to\gamma\eta)$, and $\jpsi\to\eta\ Z (\to\gamma\piz)$. 
Charge-conjugation and parity conservation allow $0^{++}$ and $2^{++}$ for $X$, and $1^{+-}$ and $1^{--}$ for $Y$ and $Z$.  
Isospin conservation further restricts $X$ and $Y$ to isospin triplets and $Z$ to an isospin singlet.

The decay amplitude for $N_{\mathrm{wave}}$ partial waves is constructed with the covariant-tensor formalism~\cite{Zou:2002ar}.
Intermediate states are described by relativistic Breit-Wigner (RBW) function~\cite{Breit:1936zzb} with corresponding PDG masses and widths, except for the $a_0(980)^0$ (Flatté~\cite{Wu:2008hx,BESIII:2021aza}) and the $\rho(770)^0$ and $\rho(1450)^0$ (Gounaris–Sakurai~\cite{Gounaris:1968mw}).
 No non-resonant contribution is included in the fit.

 The fraction of each quasi-two-body process $X$ is evaluated at both generator and reconstruction level: 
\begin{equation}
	f_{X,\mathrm{gen}} = \frac{\sum_{n=1}^{N_{\mathrm{gen}}}\mathcal{A}_X}{\sum_{n=1}^{N_{\mathrm{gen}}}\mathcal{A}},\quad
	f_{X,\mathrm{rec}} = \frac{\sum_{n=1}^{N_{\mathrm{rec}}}\mathcal{A}_X}{\sum_{n=1}^{N_{\mathrm{rec}}}\mathcal{A}},
\end{equation}
where $N_{\mathrm{gen}}$ and $N_{\mathrm{rec}}$ are the numbers of generated and reconstructed events in the PHSP MC sample (in the full $m_{\eta\piz}$ region),
respectively, and $\mathcal{A}_X$ sums over partial waves for the process $X$ only. The efficiency for process $X$ is then calculated as
$\left(f_{X,\mathrm{rec}}N_{\mathrm{rec}}\right)/\left(f_{X,\mathrm{gen}}N_{\mathrm{gen}}\right)$.

An unbinned maximum likelihood fit is performed on the data. 
To account for the signal events in the region $m_{\eta\piz}< \SI{1.0}{GeV/\clight^2}$, the object function for likelihood minimization is revised as,
\begin{align}
	 s  = &-\sum_i^{N_{\mathrm{data}}} Q_i \ln{\frac{\mathcal{A}_i}{C^+}} + \sum_i^{N_{\eta_c}}w_{\eta_c,i}\ln{\mathcal{A}_i} \notag\\
        & - \frac{\left(
	\frac{C^{-}}{C^{+}}\sum_i^{N_{\mathrm{data}}}Q_i  -N_0\right)^2}{\sigma_0^2},
	\label{eq:likelihood}
\end{align}
where 
\begin{itemize}
  \item the first term is for the signal candidates above $m_{\eta\piz}>\SI{1.0}{GeV/\clight^2}$, $N_{\mathrm{data}}$ is the number of data events, $Q_i$ is the signal weight obtained with the $Q$-weight method as described in Sec.~\ref{sec:backgroundstudy}, $C^{+}$ is the normalization factor in the $m_{\eta\piz}> \SI{1.0}{GeV/\clight^2}$ region;
  \item the second term is for the category~I background obtained with the MC simulated sample of $\jpsi\to \gamma \eta_c\to \gamma\eta \piz\piz$. $N_{\eta_c}$
		is the number of $\jpsi\to\gamma\eta_c\to\gamma\eta\piz\piz$ simulated sample, $\omega_{\eta_c, j}$ is the corresponding event weight to ensure the correct background model and the background yield \num{225\pm15}, obtained with the data-driven method as described in Sec.~\ref{sec:backgroundstudy};
  \item the third term imposes a Gaussian constraint on the signal yield $N_0=\num{676\pm80}$ in the region $m_{\eta\piz}< \SI{1.0}{GeV/\clight^2}$, which is estimated in Sec.~\ref{sec:backgroundstudy}. $C^{-}$ is the corresponding normalization factor in the region $m_{\eta\piz}< \SI{1.0}{GeV/\clight^2}$ obtained from the amplitude model.
\end{itemize}

 The amplitude model is optimized by sequentially testing all known intermediate states from the PDG compilation~\cite{10.1093/ptep/ptaa104}. 
 Each resonance is added to the baseline model individually, and only those processes exhibiting statistical significance larger than
 \num{5}$\sigma$ (evaluated from the change in likelihood and the number of additional free parameters~\cite{significance}) are retained in the nominal amplitude model. 
 For the nominal amplitude model, the yield of process $X$ is calculated as $N_X= f_{X,\mathrm{rec}}\left(\sum_i^{N_{\mathrm{data}}}Q_i+N_0\right)$, and the corresponding statistical uncertainties on the fitted yields are evaluated with 100 bootstrap samples~\cite{Efron:1979bxm,BESIII:2023exz}.

\begin{table*}[htpb]
	\centering
	\caption{Summary of statistical significance, signal yields, detection efficiencies, and $\mathrm{BF}$s of the intermediate processes in the nominal solution.
	Single uncertainties are statistical; for two uncertainties, the first is statistical and the second is systematic.} 
	\label{tab:yield}
	\begin{tabular}{lcccc}
		\hline
		Process &  Significance ($\sigma$)& Yield &  Efficiency (\si{\percent}) & $\mathrm{BF}~$($\times 10^{-6}$) \\
		\hline
		$\jpsi\to\gamma a_2(1320)^0\to\gamma\eta\piz$& \num{\gg 5} & \num{734.5\pm83.9}& \num{9.80\pm0.03} &\num{1.91\pm0.22\pm0.24}\\
		$\jpsi\to\gamma a_2(1700)^0\to\gamma\eta\piz$& \num{13.3} & \num{223.1\pm56.7}& \num{9.37\pm0.05} &\num{0.61\pm0.15\pm0.23}\\
        $\jpsi\to\gamma a_0(980)^0\to\gamma\eta\piz$& \num{8.8} & \num{137.3\pm13.6} &\num{9.42\pm0.07}&\num{0.37\pm0.04\pm0.10}\\ \hline
		$\jpsi\to\piz b_1(1235)^0\to\gamma\eta\piz$& \num{\gg 5} & \num{2861.8\pm171} & \num{10.54\pm0.02}\ \,&\num{6.92\pm0.41\pm0.63} \\
        $\jpsi\to\piz \rho(1450)^0\to\gamma\eta\piz$& \num{15.8} & \num{2256.1\pm368} & \num{11.0\pm0.02}\ \,&\num{5.23\pm0.85\pm0.83}\\
        $\jpsi\to\piz \rho(770)^0\to\gamma\eta\piz$& \num{12.1} & \num{433.7\pm82.2}&\num{7.34\pm0.03}&\num{1.50\pm0.28\pm0.51} \\ \hline
		$\jpsi\to\eta h_1(1170)\to\gamma\eta\piz$& \num{17.8} & \num{2246.9\pm248}&\num{8.33\pm0.02}&\num{6.87\pm0.76\pm1.70} \\
		$\jpsi\to\eta h_1(1595)\to\gamma\eta\piz$& \num{7.9} & \ \,\num{459.2\pm208} &\num{10.95\pm0.04}\ \,&\num{1.07\pm0.48\pm0.44}\\
		\hline
	$\jpsi\to\gamma\eta\piz$ & & \num{11137.4\pm 131}\ \ \  & \num{11.0\pm0.01}\ \,& \num{25.70\pm0.31\pm1.50}\ \,\\
		\hline
	\end{tabular}
\end{table*}

\begin{figure*}[htpb]
	\centering
	\includegraphics[width=0.8\linewidth]{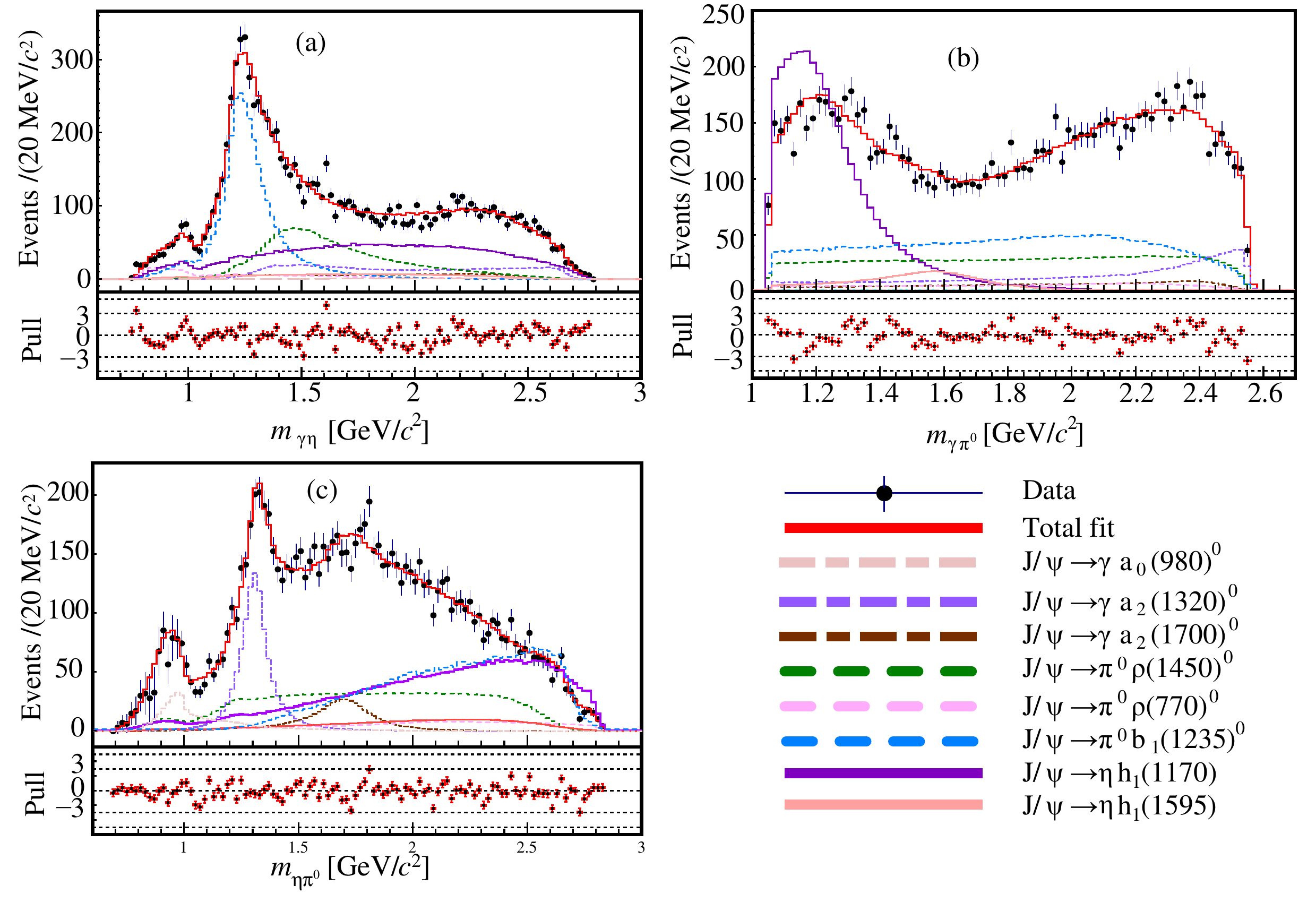}
	\caption{Projections of the nominal results on the (a) $m_{\gamma\eta}$, (b) $m_{\gamma\piz}$ and (c) $m_{\eta\piz}$ distributions. The black dots with error
	bars represent $Q$-weighted data, the red solid lines show the total fit results, and the remaining solid, dashed, and dash dotted lines correspond to the
signal contributions from individual intermediate processes, as indicated in the legend placed at the bottom right. The pull distributions
are shown in the bottom of each plot. In Fig.~(c), the background subtracted data shown in Fig.~\ref{fig:mpieta_data} is also plotted.}%
	\label{fig:projections1}
\end{figure*}

The statistical significance and yields of all intermediate processes in the nominal amplitude model are summarised in Table~\ref{tab:yield}. 
The corresponding distributions on $m_{\gamma\eta}$, $m_{\gamma\piz}$ and $m_{\eta\piz}$ are shown in Fig.~\ref{fig:projections1}, where good agreement
between data and MC projections is observed.
The masses and widths of the most significant resonances, $a_2(1320)^0$ and $b_1(1235)^0$, are allowed to float in the fit. 
The results  are consistent with the world averages~\cite{10.1093/ptep/ptaa104} except for the $b_1(1235)^0$ width, as detailed in the supplemental
material~\cite{supplementalMaterial}.
The larger fitted width of $b_1(1235)^0$ may reflect an imperfect description of the $\rho(1450)^0$ lineshape~\cite{Achasov:2013eli}, a bias in the
world average value~\cite{10.1093/ptep/ptaa104}, or the presence of a nearby axial-vector state~\cite{Clymton:2023txd}.
Larger data samples~\cite{Achasov:2023gey,*Charm-TauFactory:2013cnj}, improved input on related channels, such as $\jpsi\to
\omega\piz\piz$~\cite{BaBar:2018rkc}, and a better determination of the lineshape of the $\rho(1450)^0$ meson via the decay
$\rho(1450)^0\to\gamma\eta$~\cite{Achasov:2013eli} will be needed to clarify this issue.

All intermediate states listed by the PDG~\cite{10.1093/ptep/ptaa104}, are tested; none beyond those in the nominal model exhibit $>5\sigma$ significance. 
Furthermore, systematic scans for hypothetical $0^{++}$, $2^{++}$, $1^{--}$, and $1^{+-}$ resonances with widths of 50, 100, 200, and 300~MeV/$c^{2}$ and masses spanning the kinematically allowed region in 40~MeV/$c^{2}$ steps also reveal no significant signals. 
These results constrain possible exotic hadronic states in these decay channels, such as the recently reported $a_0(1710)$~\cite{Wang:2023aza}.

 The total BF for the decay $\jpsi\to\gamma\eta\piz$ is obtained from:
 \begin{equation}
	 \mathrm{BF}(\jpsi\to\gamma\eta\piz) = \frac{N_{\mathrm{sig}}}{N_{\jpsi}\cdot \varepsilon_{\mathrm{tot}}\cdot \mathrm{BF}_{\piz\to\gamma\gamma}\cdot \mathrm{BF}_{\eta\to\gamma\gamma}},
 \end{equation}
 where $N_{\mathrm{sig}}=N_0+\sum_i^{N_{\mathrm{data}}}Q_i$ represents the total signal yield in the full $m_{\eta\piz}$ region, $N_{\jpsi}$ is the total number
 of $\jpsi$ events in data~\cite{BESIII:2021cxx}, $\mathrm{BF}_{\piz\to\gamma\gamma}$ and $\mathrm{BF}_{\eta\to\gamma\gamma}$ are the BFs of the $\piz\to\gamma\gamma$ and $\eta\to\gamma\gamma$ decays~\cite{10.1093/ptep/ptaa104}, respectively. 
The overall detection efficiency, $\varepsilon_{\mathrm{tot}}=\SI{11.03\pm0.01}{\percent}$, is determined from the signal MC sample generated according to the nominal amplitude model, where the uncertainty comes from the MC statistics. 
 For the individual intermediate processes $X$, the BFs are calculated via  
\begin{equation}
	 \mathrm{BF}(X) = \frac{N_{X}}{N_{\jpsi}\cdot \varepsilon_{X}\cdot \mathrm{BF}_{\piz\to\gamma\gamma}\cdot \mathrm{BF}_{\eta\to\gamma\gamma}},
\end{equation}
with the process-specific efficiency $\varepsilon_X$ obtained from the PHSP signal MC sample $\varepsilon_{X}=\frac{f_{X,\mathrm{rec}}N_{\mathrm{rec}}}{f_{X,\mathrm{gen}}N_{\mathrm{gen}}}$, and $N_X$ introduced in Sec.~\ref{sec:amplitude_analysis}. 
The measured BFs are presented in Table~\ref{tab:yield}.

\section{Systematic Uncertainties}
\label{sec:systematic_uncertainties}
Systematic uncertainties are grouped into three categories.
The first category contains sources that propagate directly to the BFs results, including:
\begin{itemize}
	\item an uncertainty of \SI{0.4}{\percent} due to the total number of $\jpsi$ events~\cite{BESIII:2021cxx};
	\item an uncertainty of \SI{5}{\percent} from the photon detection, assuming \SI{1}{\percent} per photon candidate~\cite{BESIII:2016gkg};
	\item an uncertainty of \SI{0.4}{\percent} from the input BFs of the $\piz\to\gamma\gamma$ and $\eta\to\gamma\gamma$ decays~\cite{10.1093/ptep/ptaa104}.
\end{itemize}


The second category of systematic uncertainties include those associated with the other selection criteria and background treatments. These uncertainties are
estimated with alternative selection criteria and different background treatment approaches. The analysis is repeated, and the resultant deviations in the
BFs with respect to the nominal values are taken as the uncertainties. They are discussed below:

\begin{itemize}
	\item The photon energy uncertainty used in the calculation of the $\chi^2_{4C}$ is varied by \SI{4}{\percent} in the MC simulation to evaluate its impact on the kinematic fit~\cite{BESIII:2024xfs}. 
\item The $\Delta^2_{\piz}$ requirement is scanned over the range \num{0.05} to \num{0.11}~${({\rm GeV}/c^2)^2}$.  
\item The $\omega$ mass windows on $m_{\pi\gamma_{\eta}}$ and $m_{\gamma\gamma_\pi}$ are each extended by \SI{15}{MeV/\clight^2}, corresponding to $1\sigma$ of the $\omega$ mass resolution. 
\item The $\eta$ mass window on $m_{\gamma\gamma_{\eta}}$ is similarly widened by \SI{13}{MeV/\clight^2}.  
\item To evaluate the uncertainty induced by the $Q$-weight procedure, the background PDF is changed from a second-order to a third-order polynomial. In addition, the number of neighboring events used in the $Q$-weight estimation is varied from 100 and 300. 
\item To evaluate the uncertainty of the background yield in the region $m_{\eta\piz}<\SI{1.0}{GeV/\clight^2}$, an alternative fit is performed in the $\eta$
	sideband region, where the background component $\jpsi\to\gamma\eta^{\prime}\to \gamma 3\piz$ is included, and the convolved Gaussian function is removed. The resulting expected signal yield is used in the amplitude analysis.
\item Due to the small contribution from the $\jpsi\to\gamma\eta_c$ background in data, an alternative fit omitting the second term in Eq.~(\ref{eq:likelihood}) is performed. 
\end{itemize}

The third category of systematic uncertainties include those associated with the amplitude analysis model. The corresponding uncertainties are estimated with
the alternative amplitude fits, as described below.
\begin{itemize}
\item The lineshape of intermediate states is changed from a RBW function to a mass-dependent Breit-Wigner function~\cite{VonHippel:1972fg}, while keeping the resonance parameters unchanged. 
\item The fixed resonance parameters of intermediate states are varied by $\pm 1\sigma$ based on their reported uncertainties~\cite{10.1093/ptep/ptaa104}. For the $\az^0$ meson, rather than varying resonance parameters, its coupling strengths are varied within their $\pm 1\sigma$ ranges~\cite{BESIII:2021aza}.
\item Intermediate states with statistical significance between $3\sigma$ and $5\sigma$, namely $\jpsi\to\gamma a_0(1710)^0\to\gamma\eta\piz$ and $\jpsi\to\piz
	h_1(1415)\to\gamma\eta\piz$, are included in the amplitude model to estimate potential uncertainties due to additional resonances.
\end{itemize}

All uncertainties are added in quadrature to obtain the total systematic errors quoted in the supplemental material~\cite{supplementalMaterial}.

\section{Summary}%
\label{sec:summary}
The decay $\jpsi\to\gamma\eta\piz$ is studied with a data sample of \num{10087 \pm 44 e6} $\jpsi$ events collected with the \mbox{BESIII} detector. The BF is
measured to be $\mathrm{BF}(\jpsi\to\gamma\eta\piz) = \left(25.7\pm0.3\pm1.5\right)\times 10^{-6}$ excluding the processes $\jpsi\to\eta\omega(\phi)\to\gamma\eta\piz$, where the first uncertainty is statistical and the second is systematic. This result is consistent with the previous measurement~\cite{BESIII:2016gkg}, while achieving a sevenfold reduction in statistical uncertainty and an approximate twofold reduction in systematic uncertainty.

An amplitude analysis is performed for the first time on this decay, enabling the determination of the $\mathrm{BF}$s of intermediate processes, as summarized in Table~\ref{tab:yield}.
\begin{table}[h]
	\centering
	\caption{Comparison of the decay $\mathrm{BF}$ of $\jpsi\to\gamma a_0(980)^0\to\gamma\eta\piz$ measured in this work with theoretical predictions. For the measured result, the first uncertainty is statistical and the second is systematic.}
	\label{tab:a0}
	\begin{tabular}{cc}
		\hline
		~~Reference~~ & ~~$\mathrm{BF}(\jpsi\to\gamma a_0(980)^0\to\gamma\eta\piz)$ ($\times 10^{-6}$)~~ \\
		\hline
		Ref.~\cite{Sakai:2019uig} & \num{0.27} \\
		Ref.~\cite{Xiao:2019lrj} & \num{0.048} \\
		\hline
		This work & \num{0.37\pm0.04\pm0.10} \\
		\hline
	
	\end{tabular}
\end{table}

One notable result is the measurement of $\mathrm{BF}(\jpsi\to\gamma\az^0\to\gamma\eta\piz)$, which is compared with the theoretical predictions in Table~\ref{tab:a0}. 
The branching fraction for $\jpsi\to\gamma a_0(980)^0\to\gamma\eta\piz$, $(0.37\pm0.04\pm0.10)\times10^{-6}$, agrees within $0.8\sigma$ with the VMD prediction~\cite{Sakai:2019uig} but is $2.6\sigma$ above the value obtained with $a_0(980)$–$f_0(980)$ mixing dominance~\cite{Xiao:2019lrj}. 

By incorporating the measured $\mathrm{BF}$ of $\jpsi\to\piz b_1(1235)^0$ and the world average total decay width of the $b_1(1235)^0$ meson~\cite{10.1093/ptep/ptaa104}, the partial decay width of the process $b_1(1235)^0\to\gamma\eta$ is extracted. The result, presented in Table~\ref{tab:b1}, agrees with the VMD model prediction~\cite{Nagahiro:2008zza} within $0.4\sigma$, but deviates significantly (by $6.5\sigma$) from the prediction in Ref.~\cite{Lutz:2008km}.
These results support the interpretation of the $a_0(980)$ and $b_1(1235)$ mesons as dynamically generated states, with the VMD mechanism playing an important role in their radiative decays. 

\begin{table}[h]
	\centering
	\caption{Comparison of the partial decay width of $b_1(1235)^0\to\gamma\eta$ with theoretical predictions. For the measured result, the first uncertainty is statistical, the second is systematic, and the third associated with the $\mathrm{BF}$ uncertainty of $\jpsi\to\piz b_1(1235)^0$ cited from the PDG~\cite{10.1093/ptep/ptaa104}. The uncertainty from the total decay width of $b_1(1235)^0$ is negligible.}
	\label{tab:b1}
	\begin{tabular}{cc}
		\hline
		~~~~Reference~~~~ & ~~~~$\Gamma(b_1(1235)^0\to\gamma\eta)$ (\si{keV})~~~~ \\
		\hline
		Ref.~\cite{Lutz:2008km} & \num{1220} \\
		Ref.~\cite{Nagahiro:2008zza} & \num{488\pm70} \\
		\hline
		This work & \num{426\pm25.6\pm44.0\pm110.8} \\
		\hline
	\end{tabular}
\end{table}

From a broader perspective, this study offers valuable experimental inputs for ongoing efforts to understand light meson structures and interactions. While current theoretical models are consistent with several key observations, further predictions, especially those involving alternative meson configurations and decay dynamics, would increase the significance and scope of these observations.
Experimentally, the limited statistics hinder the identification of additional intermediate states in the decay $\jpsi \to \gamma \eta \piz$. 
The possible existence of exotic radiative decay channels such as $\jpsi\to\gamma a_0(1710)^0\to\gamma\eta\piz$~\cite{BESIII:2022npc} and $\jpsi\to\gamma
\pi_1(1600)^0\to\gamma\eta\piz$~\cite{E852:1997gvf}, along with the hypotheses involving two-pole structures~\cite{Clymton:2023txd,Clymton:2024pql} or unobserved members of the $b_1$ and $h_1$ meson families, remain open and compelling questions. To obtain more definitive and comprehensive results, larger $\jpsi$ data samples will be essential in future studies~\cite{Achasov:2023gey,*Charm-TauFactory:2013cnj}.

\begin{acknowledgments}

The BESIII Collaboration thanks the staff of BEPCII (https://cstr.cn/31109.02.BEPC), the IHEP computing center and the supercomputing center of the University
of Science and Technology of China (USTC) for their strong support. This work is supported in part by National Key R\&D Program of China under Contracts Nos.
2025YFA1613900, 2023YFA1606000, 2023YFA1606704, 2023YFA1609400; National Natural Science Foundation of China (NSFC) under Contracts Nos. 11635010, 11935015, 11935016, 11935018,
12025502, 12035009, 12035013, 12061131003, 12122509, 12105100, 12105276, 12192260, 12192261, 12192262, 12192263, 12192264, 12192265, 12221005, 12225509, 12235017,
12342502, 12361141819; the Chinese Academy of Sciences (CAS) Large-Scale Scientific Facility Program; the Joint Large-Scale Scientific Facility Funds of the
NSFC and CAS under Contracts No. U2032111; the CAS Youth Team Program under Contract No. YSBR-101; the Strategic Priority Research Program of Chinese Academy of
Sciences under Contract No. XDA0480600; CAS under Contract No. YSBR-101; 100 Talents Program of CAS; Beijing Natural Science Foundation (BJNSF) under Contract
No. JQ22002; The Institute of Nuclear and Particle Physics (INPAC) and Shanghai Key Laboratory for Particle Physics and Cosmology; ERC under Contract No. 758462; German Research Foundation DFG under Contract No. FOR5327; Istituto Nazionale di Fisica Nucleare, Italy; Knut and Alice Wallenberg Foundation under Contracts Nos. 2021.0174, 2021.0299, 2023.0315; Ministry of Development of Turkey under Contract No. DPT2006K-120470; National Research Foundation of Korea under Contract No. NRF-2022R1A2C1092335; National Science and Technology fund of Mongolia; Polish National Science Centre under Contract No. 2024/53/B/ST2/00975; STFC (United Kingdom); Swedish Research Council under Contract No. 2019.04595; U. S. Department of Energy under Contract No. DE-FG02-05ER41374
\end{acknowledgments}

%

\end{document}